\documentclass[twocolumn,showpacs,aps,amsmath,amssymb,floatfix,superscriptaddress]{revtex4}
\pdfoutput=1 
\usepackage{graphicx,eucal}

\begin{document}
\title{Reaction-Diffusion Processes with Nonlinear Diffusion}
\author{P. L. Krapivsky}
\affiliation{Department of Physics, Boston University, Boston, Massachusetts 02215, USA}

\begin{abstract}
We study reaction-diffusion processes with concentration-dependent diffusivity. First, the decay of the concentration in the single-species and two-species diffusion-controlled annihilation processes is determined. We then consider two natural inhomogeneous realizations. The two-species annihilation process is investigated in the situation when the reactants are initially separated, namely each species occupies a half space. In particular, we establish the growth law of the width of the reaction zone. The single-species annihilation process is studied in the situation when the spatially localized source drives the system toward the non-equilibrium steady state. Finally, we investigate a dissolution process with a localized source of diffusing atoms which react with the initially present immobile atoms forming immobile molecules. 
\end{abstract}
\maketitle

\section{Introduction}

Random walks and their continuum descriptions in terms of convection-diffusion equations underlie numerous natural phenomena \cite{B93,Hughes,V01}. Our experience with these equations strongly influences our intuition. For instance, an instantaneous propagation of perturbations is a key property of the diffusion equation which appears puzzling when we first learn it; eventually, we start to regard it as a general feature of parabolic partial differential equations (PDEs) which distinguishes them from hyperbolic PDEs. It then comes as a surprise that the instantaneous propagation of perturbations does not generally apply to nonlinear parabolic PDEs, so that phenomenologically the predicted behaviors resemble the behaviors of hyperbolic PDEs. Perhaps the simplest class of such nonlinear parabolic PDEs is 
\begin{equation}
\label{NLDE:general}
\frac{\partial c}{\partial t} = \nabla\cdot \left(c^a\,\nabla c\right)
\end{equation}
According to \eqref{NLDE:general}, the spread of a localized initial profile into a vacuum [that is, $c({\bf r}, t=0)=0$ when $r=|{\bf r}|>R$ for some finite $R$] is {\em not} instantaneous whenever $a>0$. More precisely, the front advances as $r_f\sim t^{1/(2+da)}$ (here $d$ is the spatial dimension); for $r>r_f$, the medium is in the vacuum state \cite{Zeld,LL87,B96,book,Evans}. Of course, there is no real contradiction with the standard lore as Eq.~\eqref{NLDE:general} is {\em nonlinear}. This spectacular effect occurs when we consider the spread into a vacuum (where the diffusion coefficient $D=c^a$ vanishes). 

Throughout this article we shall tacitly assume that  $D(c)=0$ only when $c=0$. One should keep in mind, however, that  for some lattice gases the diffusion coefficient vanishes at a certain positive critical density \cite{LS84,Spohn}. Near the critical density, a power law asymptotic, $D\sim |c-c_*|^a$, usually holds \cite{Spohn}.  Hence if initially $c\geq c_*$, we recover \eqref{NLDE:general} by making a shift, $c\to c-c_*$. 

Nonlinear parabolic PDEs form a fertile research area, the behavior of solutions to nonlinear parabolic PDEs is still poorly understood \cite{Evans}. We emphasize that  nonlinear parabolic PDEs similar to Eq.~\eqref{NLDE:general} and their microscopic brethren are by no means pathological. For instance, Navier-Stokes equations are nonlinear parabolic PDEs with temperature-dependent transport coefficients. Thus in describing heat conduction in the simplest case when the fluid velocity vanishes (heat conduction without convection) we must solve a nonlinear parabolic PDE for the temperature $T$ similar to Eq.~\eqref{NLDE:general} as the coefficient of thermal conductivity $\chi$ depends on the temperature \cite{Zeld,LL87}. For instance, for the hard-sphere gas $\chi\propto \sqrt{T}$ indicating that heat conduction in the hard-sphere gas is described by Eq.~\eqref{NLDE:general} with $a=\frac{1}{2}$. The concentration-dependent diffusion effect has been observed in ionic crystals and oxides \cite{MF78,S82,W84,E89}, it is relevant in electrochemistry \cite{BR02,HCCP08}, and it has  many other applications \cite{KL04}. The concentration-dependent diffusion coefficients have been also derived in the realm of some microscopic models \cite{LB_08}. 

Nonlinear parabolic PDEs often arise as mathematical models of reaction-diffusion processes \cite{book}. Almost all studies of reaction-diffusion processes rely on the standard diffusive transport. Reaction-diffusion processes with density-dependent hopping rates have appeared in a few studies \cite{Rosenau,BLL11}, e.g., annihilation processes have been investigated in Ref.~\cite{BLL11}. Here we examine basic reaction-diffusion processes in the situation when the transport is described by \eqref{NLDE:general}. 

The rest of this article is organized as follows. In Secs.~\ref{AP}--\ref{AP2} we study diffusion-controlled single-species and two-species annihilation processes, respectively. We examine the spatially homogeneous setting when particles are distributed at random with uniform concentration for the single-species annihilation process and with uniform and equal concentrations for the two-species annihilation process. In Sec.~\ref{IN_AP}, we consider the two-species annihilation process in the situation when the reactants are initially spatially separated, viz. two half spaces are occupied by dissimilar species. In Sec.~\ref{source}, we analyze the single-species annihilation process driven by a localized steady source. In Sec.~\ref{DP}, we study a dissolution process.  This process involves three distinct species, viz. diffusing atoms which are injected into a localized region and react with immobile atoms creating immobile molecules. We summarize our results in Sec.~\ref{Sum}. 

\section{Single-species annihilation process}
\label{AP}

One of the simplest reaction-diffusion processes is the diffusion-controlled single-species annihilation. Symbolically, this process is represented by the reaction scheme
\begin{equation}
\label{annih}
A+A\to \emptyset
\end{equation}
where identical diffusing particles (say atoms) are denoted by $A$. The process \eqref{annih} postulates that when the two particles collide, they disappear. A true annihilation is occasionally possible (e.g., the annihilation of domain walls in spin chains). In most situations, however, a collision between two atoms will lead to the formation of a diatomic molecule; if such molecules are stable (that is, they do not break back to atoms) and if molecules do not influence diffusing atoms, we can effectively ignore the molecules and use the reaction scheme \eqref{annih}. 

The diffusion-controlled single-species annihilation process \eqref{annih} with concentration-independent diffusivity is described by the reaction-diffusion equation
\begin{equation}
\label{RD_simple}
\frac{\partial c}{\partial t} = D\nabla^2 c - Kc^2
\end{equation}
The reaction rate $K$ depends on the details of the process; in the simplest case when spherical particles of equal radii $R$ diffuse independently and immediately react upon colliding, the reaction rate theory expresses $K$ through the diffusion constant $D$ and $R$, viz.   
\begin{equation}
K = 16\pi D R
\end{equation}
This well-known Smoluchowski formula \cite{SM,C43,OTB89} is valid in three dimensions. More generally, in $d$ dimensions the reaction rate scales according to $K \sim DR^{d-2}$  \cite{book}. In the spatially homogeneous setting, $\dot c = -Kc^2$, leading to the $c\simeq (Kt)^{-1}$ large time decay of the concentration. This is valid in $d>2$ dimensions; for $d\leq 2$, the above formula predicts $c\sim R^{2-d}/(Dt)$, so the $R-$dependence is certainly wrong when $d\leq 2$.  A heuristic way of understanding the behavior in low dimensions relies on the basic properties of random walks, particularly on the estimates for the average number of distinct sites visited by a random walker \cite{Hughes}. In the spatially homogeneous setting, one gets 
\begin{equation}
\label{RD_hom}
\frac{dc}{dt}\sim -
\begin{cases}
DR c^2                                                 & d=3\\
D c^2\,  \big[\ln\frac{1}{cR^2}\big]^{-1} & d=2\\
Dc^3                                                     & d=1
\end{cases}
\end{equation}
from which we deduce the well-known asymptotic behaviors 
\begin{equation}
\label{DLA}
c\sim 
\begin{cases}
t^{-1}       & d=3\\
t^{-1}\ln t & d=2\\
t^{-1/2}    & d=1
\end{cases}
\end{equation}
In equation \eqref{DLA} we have displayed only time dependence; using parameters $D$ and $R$ one easily restores the dimensionally correct behavior [e.g., in two dimensions $c\sim (Dt)^{-1}\ln(Dt/R^2)$]. The asymptotic behaviors \eqref{DLA} have been established through a combination of simulations, heuristic arguments (as outlined above), and exact solutions in one dimension; they have been subsequently proven in Refs.~\cite{BG80,A81}. 

The underlying microscopic process is simple to realize on a lattice. Each site is occupied by at most one particle, and if a particle hops to the occupied site, both particles annihilate. Particles undergo symmetric nearest-neighbor hopping. When the hopping rates are constant, the density decays according to Eq.~\eqref{DLA}.

The simplest way to mimic the $D=c^a$ hopping rate is to postulate that the current hopping rate of each particle is inversely proportional to the distance $\ell$ between the particle and its nearest neighbor, namely
\begin{equation}
\label{hopping}
\text{hopping rate} = \ell^{-a/d}
\end{equation}
 The quantity $\ell^{-1/d}$ can be interpreted as a local concentration, and hence the hopping rule \eqref{hopping} mimics the macroscopic $D\sim c^a$ dependence. 
 
For the single-species annihilation process \eqref{annih} with concentration-dependent hopping rate the generalization of \eqref{RD_hom} is simple: one merely replaces the constant $D$ by the concentration-dependent $D=c^a$. This yields 
\begin{equation}
\label{RD_non}
\frac{dc}{dt}\sim -
\begin{cases}
c^{2+a} & d=3\\
c^{2+a}\,\big[\ln\frac{1}{cR^2}\big]^{-1} & d=2\\
c^{3+a}     & d=1
\end{cases}
\end{equation}

A heuristic derivation of \eqref{RD_non} is equivalent to the derivation of \eqref{RD_hom}. In three dimensions, one writes $\dot c=-Kc^2$, where the factor $c^2$ merely reflects the binary nature of the annihilation process; then the reaction rate theory (see Refs.~\cite{book,OTB89}) gives $K\sim DR\propto c^a$ thereby leading to $\dot c\sim -c^{2+a}$ as stated in \eqref{RD_non}. In one and two dimensions, the recurrent nature of the random walks becomes crucial. Generally we write $\dot c=-c/\tau$, so we need to estimate  the collision time $\tau$. Such estimates follow from the known expressions for the average number of distinct sites visited by a random walker \cite{Hughes}. In one dimension we use $\ell\sim\sqrt{D\tau}$ where $\ell\sim c^{-1}$ is the typical separation between the nearest particles. In two dimensions we need a slightly more complicated formula $D\tau\big[\ln\frac{D\tau}{R^2}\big]^{-1}\sim c^{-1}$. The asymptotic behavior of the solutions to Eq.~\eqref{RD_non} is
\begin{equation}
\label{DLA_non}
c\sim 
\begin{cases}
t^{-1/(1+a)}                & d=3\\
t^{-1/(1+a)}\, [\ln t]^{1/(1+a)}  & d=2\\
t^{-1/(2+a)}                & d=1
\end{cases}
\end{equation}

For diffusion-controlled single-species annihilation processes with concentration-independent diffusivity there is a large body of knowledge (ranging from exact solutions in one dimension to proofs in all dimensions \cite{BG80,A81}) corroborating the asymptotic behaviors \eqref{DLA} and similar results for more complicated processes including aggregation, two-species annihilation, etc., see \cite{S88,T89,B78,ZO78,TW83,KR85,BL88}. This supports the validity of \eqref{DLA_non} since the above (admittedly heuristic) arguments directly extend the well-established results \eqref{RD_hom}--\eqref{DLA} to the concentration-dependent case.  

The generalization of rigorous work \cite{BG80,A81} to the concentration-dependent hopping rates appears feasible, although it may require substantial effort. There is little hope of solving the model (even in one dimension in the simplest case of $a=1$). Finally, we note that for a ferromagnetic Ising spin chain supplemented with (conservative) spin-exchange Kawasaki dynamics, the low-temperature behavior is well represented in terms of domain walls undergoing single-species annihilation with a diffusion coefficient proportional to their density \cite{Cor_Kaw,Sat_Kaw,BK_Kaw}. This corresponds to $a=1$ and therefore according to Eq.~\eqref{DLA_non} the density of the domain walls should decay as $t^{-1/3}$ in one dimension. The celebrated $t^{-1/3}$ decay law is very well confirmed, both generally for conservative dynamics and specifically for the Kawasaki dynamics \cite{Cor_Kaw,Sat_Kaw,BK_Kaw}. 

\section{Two-species annihilation process}
\label{AP2}

\begin{figure}
\centerline{
\includegraphics[width=8cm]{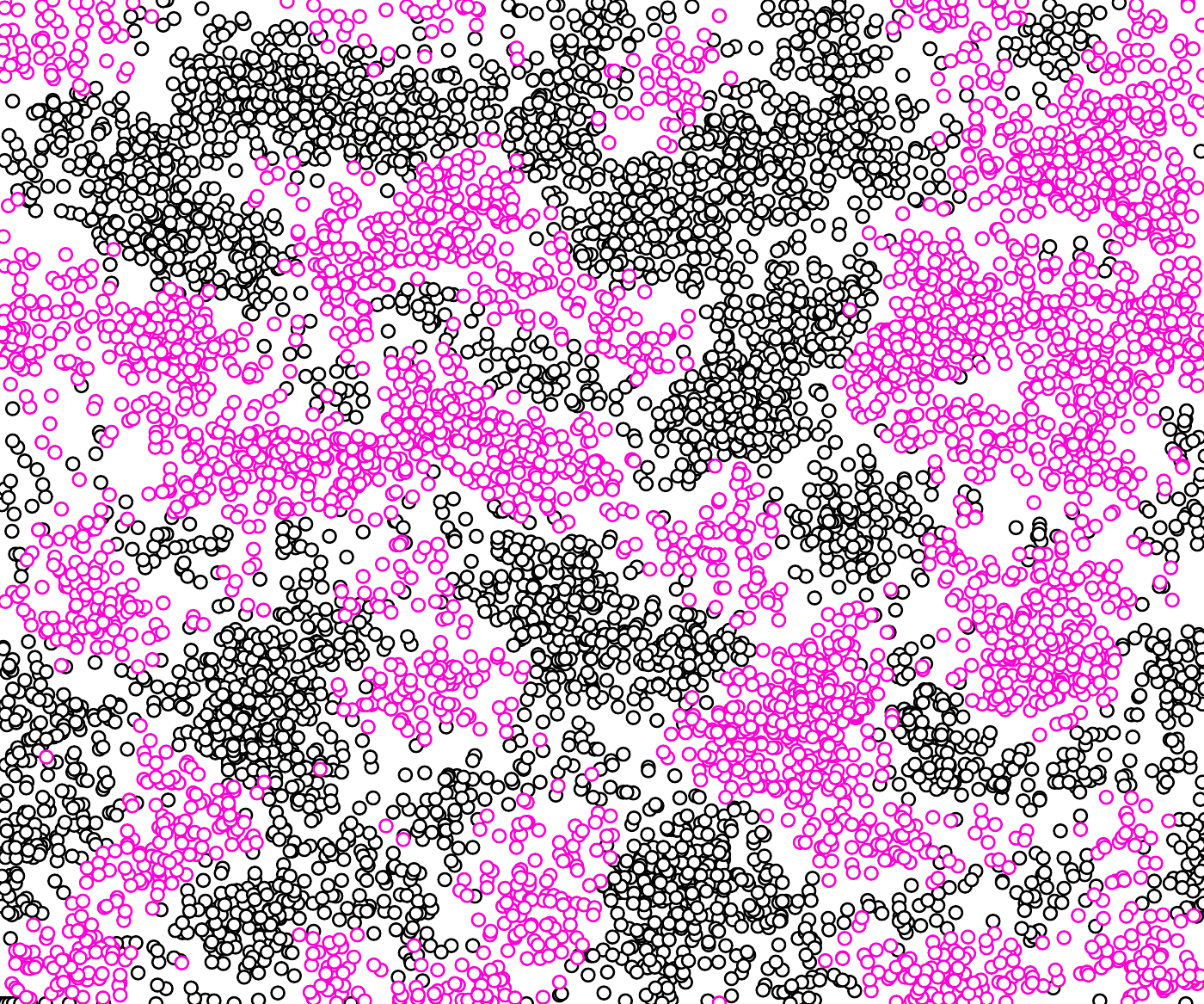}}
\caption{Snapshot of the particle positions in two-species annihilation in two dimensions (the particle radii are enlarged for visibility).}
\label{mosaic}
\end{figure}

The two-species annihilation process is represented by the reaction scheme
\begin{equation}
\label{annih_2}
A+B\to \emptyset
\end{equation}
Here $A$ and $B$ denote diffusing particles of different types and the process \eqref{annih_2} postulates that when the two dissimilar particles occupy the same lattice site, they immediately annihilate. The interesting situation arises when the initial densities are equal. (Otherwise, the minority species quickly disappears.) The critical dimension for this two-species annihilation process is not affected by the dependence of the hopping rates on concentration, and therefore $d_c=4$ as in the case of constant hopping rates \cite{B78,ZO78,TW83,KR85,BL88}. 

For $d\geq d_c$, the mean-field description is applicable: $\dot c\sim -DRc^2\sim -c^{2+a}$, and we recover the $c\sim t^{-1/(1+a)}$ decay. Below the critical dimension, the densities decay slower with time, namely as $(Dt)^{-d/4}$. This slow kinetics arises because opposite-species reactants organize into a coarsening domain mosaic (Fig.~\ref{mosaic}) and annihilation can occur only along domain boundaries rather than throughout the system \cite{B78,ZO78,TW83,KR85,BL88}. A heuristic explanation of the density decay is well known  and has already appeared in textbooks \cite{book}. This explanation is based on the domain picture. Namely, one argues that in a domain of size $L\sim \sqrt{Dt}$ one species was in the majority (due to fluctuations in the particle numbers), so the surplus survives and since it scales as $\sqrt{c_0 L^d}$, the resulting density is $c\sim L^{-d} \sqrt{c_0 L^d}\sim \sqrt{c_0}\,(Dt)^{-d/4}$.   

Replacing the hopping rate by its average value, $D=c^a$, we obtain $c\sim \sqrt{c_0}\,(c^at)^{-d/4}$ for $d<4$. Therefore $c\sim t^{-1/(a+4/d)}$, while the mean-field $c\sim t^{-1/(1+a)}$ decay is restored when $d>d_c=4$. 

In the physically relevant dimensions, the asymptotic decay of the concentration is therefore
\begin{equation}
\label{DLA_2}
c\sim 
\begin{cases}
c_0^{2/(3a+4)}\, t^{-1/(a+4/3)}             & d=3\\
c_0^{1/(a+2)}\,  t^{-1/(a+2)}                 & d=2\\
c_0^{2/(a+4)}\, t^{-1/(a+4)}                  & d=1
\end{cases}
\end{equation}

\begin{figure}
\centerline{
\includegraphics[width=8cm]{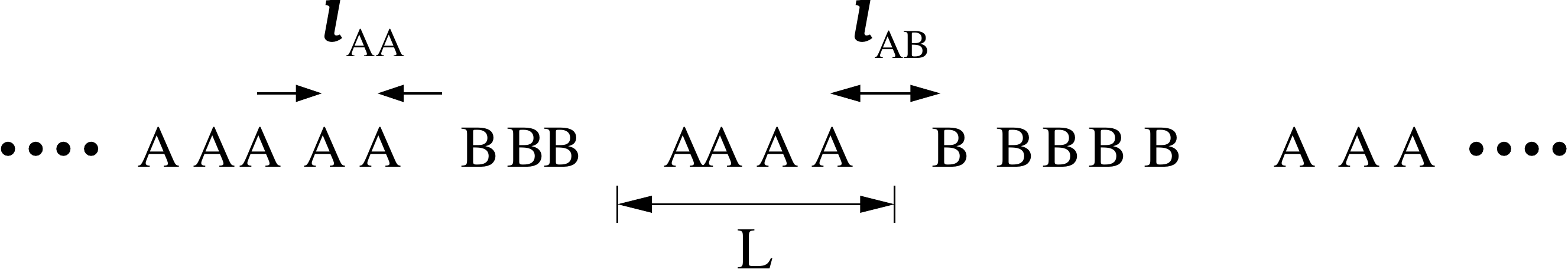}}
\caption{Illustration of the three length scales in two-species annihilation in one dimension.}
\label{scales}
\end{figure}

Interestingly, in one and two spatial dimensions the coarsening domain mosaic (Fig.~\ref{mosaic}) is characterized by three length scales \cite{LR92}. This is particularly obvious in one dimension (Fig.~\ref{scales}) where one can identify the domain size $L\sim \sqrt{Dt}$, the average spacing between adjacent particles in the same domain $\ell_{AA}=\ell_{BB}\sim c^{-1}$, and the depletion zone between the domains. This last quantity scales as $\ell_{AB}\sim c_0^{-1/4}(Dt)^{3/8}$, see \cite{LR92} or \cite{book}. In the present case of the concentration-dependent hopping rate the three length scales characterizing the two-species annihilation process in one dimension are 
\begin{equation}
\label{lengths_1d}
\begin{split}
&L\sim c_0^{\frac{a}{a+4}}\,t^{\frac{2}{a+4}}\\
&\ell_{AB}\sim c_0^{-\frac{2-a}{2(a+4)}}\,t^{\frac{3}{2(a+4)}}\\
&\ell_{AA}=\ell_{BB}\sim c_0^{-\frac{2}{a+4}}\,t^{\frac{1}{a+4}}
\end{split}
\end{equation}

Similarly generalizing the two-dimensional result  \cite{LR92} one gets
\begin{equation}
\label{lengths_2d}
L\propto t^{\frac{1}{a+2}}\,, 
\quad \ell_{AB}\propto t^{\frac{2}{3(a+2)}}\,, 
\quad \ell_{AA}=\ell_{BB}\propto t^{\frac{1}{2(a+2)}}
\end{equation}
Numerical verifications of these results could be very challenging even in one dimension. Indeed, if the hopping rate is proportional to the concentration ($a=1$), the length scales \eqref{lengths_1d} become $L\propto t^{2/5}, ~\ell_{AB}\propto t^{3/10}$ and  $\ell_{AA}\propto t^{1/5}$, so the exponents differ only by 0.1 which is hard to accurately measure. 

\section{Inhomogeneous two-species annihilation}
\label{IN_AP}

\begin{figure}
\centerline{
\includegraphics[width=8cm]{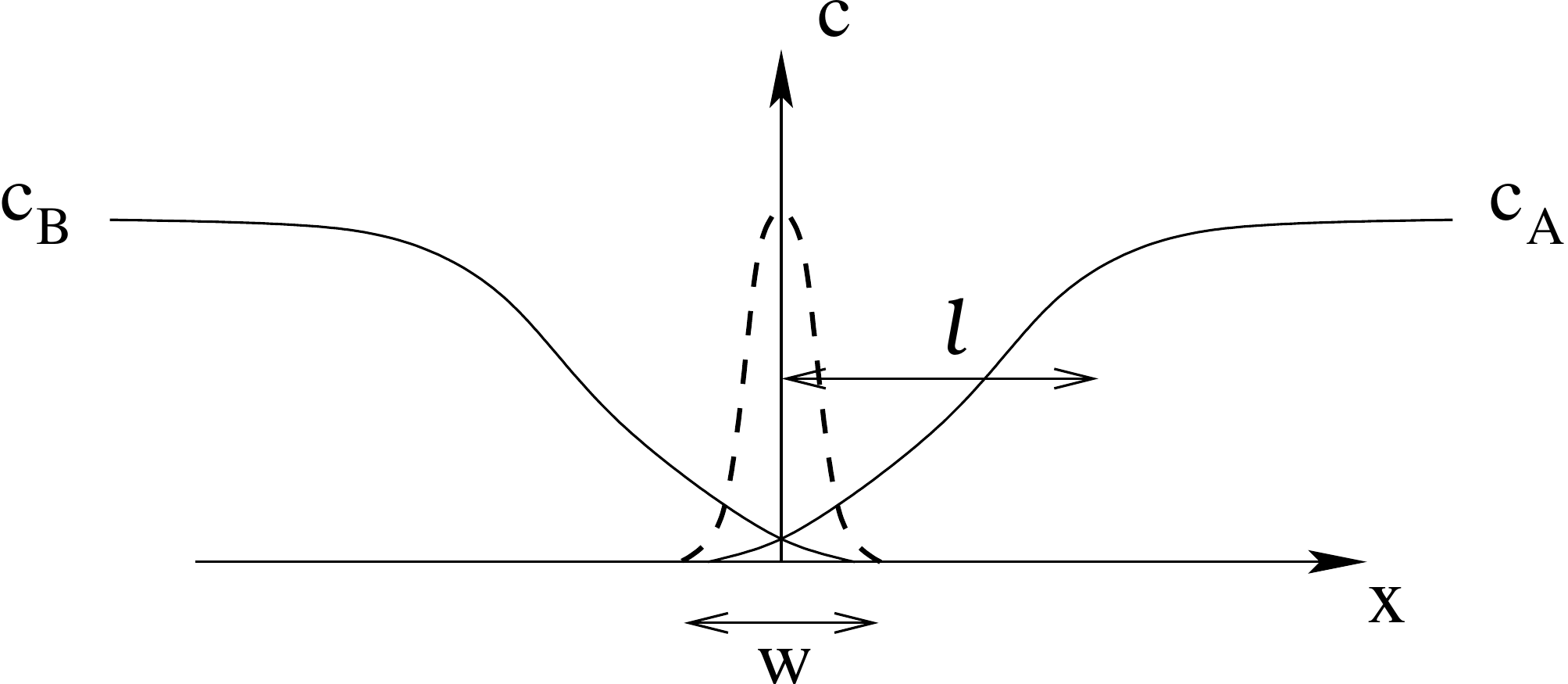}}
    \caption{Sketch of the concentration profiles (solid curves) and the
      reaction rate $Kc_Ac_B$ (dashed curve, and considerably magnified) in
      two-species annihilation with initially separated components.  The
      depletion zone width is $\ell\sim \sqrt{t}$, while the reaction zone
      width is $w\sim t^{1/(4a+6)}$.}
\label{fig-react-gr}
\end{figure}

In this section we return to the two-species annihilation process from Sec.~\ref{AP2}, but instead of uniform initial densities we assume that the reactants are initially spatially separated.  Specifically, we assume that the right (left) half space is initially occupied by $A$ ($B$) particles (Fig.~\ref{fig-react-gr}). Physically it may be realized by having a membrane at $x=0$ separating the reactants, removing the membrane at time $t=0$, and observing the subsequent reaction kinetics. The governing reaction-diffusion equations read
\begin{subequations}
\begin{align}
&\frac{\partial c_A}{\partial t} = \frac{\partial}{\partial x}\Big(D\frac{\partial c_A}{\partial x}\Big) -K c_A c_B
\label{react-A}\\
&\frac{\partial c_B}{\partial t} =  \frac{\partial}{\partial x}\Big(D\frac{\partial c_B}{\partial x}\Big)-K c_A c_B
\label{react-B}
\end{align}
\end{subequations}
where $c_A=c_A(x,t)$ and $c_B=c_B(x,t)$ denote the concentration of each species at position $x$ at time $t$. We assume that the diffusion coefficients of both species are equal and depend on the total concentration  $c_A+c_B$. 

The initial densities are
\begin{equation*}
c_A(x, t=0) =
\begin{cases}
c_0,  &x>0;\\
0,      &x<0;
\end{cases}
\end{equation*}
and $c_B(x,t=0)=c_A(-x, t=0)$.  In the constant-diffusion case, this problem has been originally investigated in Refs.~\cite{Z49,GR86}; various generalizations have been treated later, see \cite{BR92,CD93,LC94,K95,Koza96,KSPS08,MDR09} and references therein. 

We apply the same reasoning to the concentration-dependent case. Subtracting \eqref{react-B} from \eqref{react-A} we find that $c=c_A-c_B$ satisfies a diffusion equation
\begin{equation}
\label{DE:excess}
\frac{\partial c}{\partial t} = \frac{\partial}{\partial x}\Big(D\frac{\partial c}{\partial x}\Big)
\end{equation}
The domain of one species acts as a nearly fixed absorbing boundary condition for the opposite species. Indeed, the ratio $w/\ell$ of the width of the reaction zone $w$ to the width of the depletion zone $\ell$ asymptotically approaches to zero as we shall see below, and therefore in the analysis of Eq.~\eqref{DE:excess} we can treat the reaction zone as an interface and summarize its influence by using the absorbing boundary condition. Thus the density profile of each particle species, namely $A$ particles in the half space $x>0$ and $B$ particles in the half space $x<0$, is close to that of independent diffusing particles in the presence of an absorbing boundary at $x=0$. The diffusion coefficient depends on both concentrations, $D=(c_A+c_B)^a$, so \eqref{DE:excess} is not a closed equation. Away from the reaction zone, however, one of the species dominates. Thus for $x\gg w$ we can re-write \eqref{DE:excess} as
\begin{equation}
\label{excess}
c_t = (c^a c_x)_x
\end{equation}
We want to solve \eqref{excess} in the half space $x>0$ subject to the initial condition
\begin{equation}
\label{excess:IC}
c(x>0, t=0) = c_0
\end{equation}
and the absorbing boundary condition
\begin{equation}
\label{excess:BC}
c(x=0, t>0) = 0
\end{equation}
The problem \eqref{excess}--\eqref{excess:BC} admits a self-similar solution \cite{diff_caveat}
\begin{equation}
\label{cF:scaling}
c(x,t) = c_0 F(\xi), \quad\xi = \frac{x}{\sqrt{c_0^a t}}
\end{equation}
The scaling function $F(\xi)$ is a solution of the differential equation
\begin{equation}
\label{F:eq}
(F^a F')' + \tfrac{1}{2}\xi F' = 0
\end{equation}
subject to the boundary conditions
\begin{equation}
\label{F:BC}
F(0) = 0, \quad F(\infty) = 1
\end{equation}
The boundary-value problem \eqref{F:eq}--\eqref{F:BC} does not admit an analytical solution, but we are interested in the small $\xi$ behavior which can be extracted analytically (up to numerical factors). Indeed, an asymptotic analysis gives
\begin{equation}
\label{F:asymp}
F\sim \xi^{1/(a+1)}\qquad\text{when}\quad \xi\to 0
\end{equation}

We now estimate the width $w$ of the reaction zone by balancing the diffusive flux into this zone with the total rate at which the particles are annihilated. The flux is given by $2|c^a c_x|$ and using Eqs.~\eqref{cF:scaling} and \eqref{F:asymp} we find that the flux scales as $c_0^{a+1}/\sqrt{c_0^at}$.  Using the reaction-diffusion equations \eqref{react-A}--\eqref{react-B}, the number of reactions per unit time equals the integral of $Kc_Ac_B$ over the extent of the reaction zone.  We estimate this integral as $c^{a+2}w$ (since $K\sim c^a$) and use $c=c(x\!=\!w,t)$ as the typical concentration. We simplify $c^{a+2}w$ using Eqs.~\eqref{cF:scaling} and \eqref{F:asymp} and equate the result with the flux to yield
\begin{equation*}
wc_0^{a+2}\left(\frac{w}{\sqrt{c_0^at}}\right)^{\frac{a+2}{a+1}} \sim \frac{c_0^{a+1}}{\sqrt{c_0^at}}
\end{equation*}
{}from which
\begin{equation}
\label{width_3d}
w \sim c_0^{-(a+2)/(4a+6)}\, t^{1/(4a+6)} 
\end{equation}
We see that the width of the reaction reaction zone exhibits a rather slow growth $w\sim t^{1/(4a+6)}$, while the width of the depletion zone grows diffusively, $\ell\sim \sqrt{t}$. Hence the ratio $w/\ell \sim t^{-(a+1)/(2a+3)}$ asymptotically vanishes and this provides an a posteriori justification of the usage of the absorbing boundary condition \eqref{excess:BC}. 

The above analysis applies when the annihilation process is happening in three (or higher) dimensions. We now show how to handle the one- and two-dimensional cases. (See Refs.~\cite{CD93,LC94,K95} for the analysis of such low-dimensional settings in the case of constant diffusion coefficient.) In these situations, we can use \eqref{excess}--\eqref{F:asymp}, so the flux still scales as $c_0^{a+1}/\sqrt{c_0^at}$.  The reaction rate $c^{a+2}$ should be replaced by $c^{a+3}$ in one dimension, and by $c^{a+2}\big[\ln\frac{1}{cR^2}\big]^{-1}$ in two dimensions, see Eq.~\eqref{RD_non}.  Using these results we estimate the total rate at which the particles are annihilated and balance the corresponding estimates with the flux. This yields
\begin{equation}
\label{width_1d}
w \sim c_0^{-1/2}\, t^{1/(2a+4)} 
\end{equation}
in one dimension and
\begin{equation}
\label{width_2d}
w \sim c_0^{-(a+2)/(4a+6)}\, t^{1/(4a+6)}\,(\ln t)^{(a+1)/(2a+3)}
\end{equation}
in two dimensions. Combining \eqref{width_3d}--\eqref{width_2d} and ignoring the dependence on the initial concentration we conclude that the width of the reaction reaction zone grows as
\begin{equation}
\label{width}
w\sim 
\begin{cases}
t^{1/(4a+6)}                                    & d=3\\
t^{1/(4a+6)}\,(\ln t)^{(a+1)/(2a+3)} & d=2\\
t^{1/(2a+4)}                                   & d=1
\end{cases}
\end{equation}

\section{Annihilation Process Driven by a Localized Source}
\label{source}

We now return to the single-species annihilation process and consider the system which is initially empty and is driven by a localized source, say at the origin, which is turned on at time $t=0$. In the case of classical (concentration-independent) diffusion, the qualitative behaviors are largely understood \cite{CRL89,K94}; we now review the approach and then apply it to non-linear diffusion. 

\subsection{Constant Diffusion}

The corresponding reaction-diffusion equation reads 
\begin{equation}
\label{RD_source}
\frac{\partial c}{\partial t} = D\nabla^2 c - Kc^2 + J\,\delta({\bf r})\,\theta(t)
\end{equation}
Here $D$ is the diffusion coefficient which is assumed to be constant in this subsection, $J$ the strength of the source, and $\theta(t)$ the Heaviside step function assuring that the source is turned on at $t=0$. Equation \eqref{RD_source} is applicable when $d>2$; for $d=2$ and $d=1$, the reaction term must be modified according to the right-hand side of \eqref{RD_hom}.  

Assuming the emergence of the stationary concentration profile one gets
\begin{equation}
\label{RD_source_steady}
0 = D\nabla^2 c - Kc^2 + J\,\delta({\bf r})
\end{equation}
Balancing the first two terms, $D\nabla^2 c = Kc^2$, leads to 
\begin{equation}
\label{below}
c = (4-d)\,\frac{2D}{K}\,\frac{1}{r^2}
\end{equation}
Thus we see the emergence of the upper critical dimension $d^c=4$. Indeed, the result of Eq.~\eqref{below} clearly holds only when $d<d^c=4$. For $d>d^c=4$ the reaction term becomes irrelevant far away from the source. Hence one must solve the Laplace equation $D\nabla^2 c + J\,\delta({\bf r}) = 0$. The solution is 
\begin{equation}
\label{above}
c \sim \frac{J}{D}\,\frac{1}{r^{d-2}}
\end{equation}
At $d=d^c$ it is natural to expect logarithmic corrections. Indeed, one finds \cite{book,CRL89,K94}
\begin{equation}
\label{at}
c = \frac{2D}{K}\,\frac{1}{r^2\,\ln r}
\end{equation}
The behavior \eqref{below} is actually valid as long as the dimension exceeds the critical $d_c=2$. 
At $d=d_c$ we must use a modified reaction term, so we balance $\nabla^2 c \sim c^2/\ln(1/c)$ to yield
$c\sim r^{-2}\ln r$. For $d=1$ we balance $\nabla^2 c \sim c^3$ and get $c\sim r^{-1}$. Combing these results with \eqref{below}--\eqref{at} we arrive at \cite{book,CRL89,K94}
\begin{equation}
\label{c_steady}
c(r)\sim 
\begin{cases}
r^{-(d-2)}           & d>4\\
r^{-2}(\ln r)^{-1} & d=4\\
r^{-2}                 & d=3\\
r^{-2} \ln r          & d=2\\
r^{-1}                 & d=1
\end{cases}
\end{equation}

It is interesting to estimate the asymptotic growth of the total number of particles in the system. In principle,
\begin{equation*}
N(t) = \int_0^\infty dr \,\Omega_d r^{d-1} c(r,t)
\end{equation*}
where $\Omega_d$ is the surface area of unit sphere in the $d$ dimensional space: $\Omega_1=2$, $\Omega_2=2\pi$,  $\Omega_3=4\pi$, etc. 

After the source is turned on at time $t=0$, the concentration advances diffusively, that is, as $\sqrt{Dt}$. Inside the sphere of radius 
\begin{equation}
\label{front_diff}
\mathcal{R}(t)\sim \sqrt{Dt}
\end{equation}
the concentration is close to the steady-state distribution \eqref{c_steady}, while the region outside this sphere is essentially empty. Therefore
\begin{equation}
\label{estimate}
N(t) \sim \int_0^{\mathcal{R}(t)} dr \,\Omega_d r^{d-1} c(r)
\end{equation}
Inserting \eqref{c_steady} and \eqref{front_diff} into \eqref{estimate} one finds \cite{book,CRL89,K94}
\begin{equation}
\label{total}
N(t)\sim 
\begin{cases}
t                 & d>4\\
t(\ln t)^{-1} & d=4\\
t^{1/2}        & d=3\\
(\ln t)^2      & d=2\\
\ln t            & d=1
\end{cases}
\end{equation}
For $d>4$, there is so much room that particles do not `see' each other once they are sufficiently far from the origin. Overall the finite fraction of particles survive; the total number of surviving particles is of course smaller than the total number of introduced particles $Jt$, but it still grows linearly. For $d\leq 4$, the vanishing fraction of the introduced particles survives. 

\subsection{Concentration-Dependent Diffusion}

In the case of concentration-dependent diffusion, the governing equation reads 
\begin{equation}
\label{non_source}
\frac{\partial c}{\partial t} = \nabla\cdot \left(c^a\,\nabla c\right) - c^{2+a} + J\,\delta({\bf r})\,\theta(t)
\end{equation}
when $d>2$, while for $d=2$ and $d=1$ the reaction term must be modified according to the right-hand side of \eqref{RD_non}.  We again assume the emergence of the stationary concentration profile. 
Balancing the reaction and diffusion terms $\nabla\cdot \left(c^a\,\nabla c\right) = c^{2+a}$ one finds 
$c=2(4+2a-d)r^{-2}$ which tells us that the upper critical dimension is now given by $d^c=4+2a$ and the above expression for the steady-state concentration is valid when $2<d<d^c$. Repeating the same arguments as above we determine the concentration in every dimension to yield  
\begin{equation}
\label{c_steady_non}
c(r)\sim 
\begin{cases}
r^{-(d-2)/(1+a)}           & d>d^c=4+2a\\
r^{-2}(\ln r)^{-1} & d=d^c\\
r^{-2}                 & 2<d<d^c\\
r^{-2} \ln r          & d=2\\
r^{-1}                 & d=1
\end{cases}
\end{equation}

In the classical case, the position of the front \eqref{front_diff} was universal (independent on the spatial dimensionality). Now the situation is different. To determine the time dependence of the front position we use the chief formula $\mathcal{R}\sim \sqrt{Dt}$ together with $D\sim [c(\mathcal{R})]^a$. Thus we have $\mathcal{R}^2\sim [c(\mathcal{R})]^a \, t$, and using \eqref{c_steady_non} to estimate $c(\mathcal{R})$ we conclude that 
\begin{equation}
\label{front_non}
\mathcal{R}(t)\sim 
\begin{cases}
t^{(1+a)/(2+da)}                          & d>d^c\\
t^{1/(2+2a)} (\ln t)^{-a/(2+2a)}    & d=d^c\\
t^{1/(2+2a)}                                & 2<d<d^c\\
t^{1/(2+2a)} (\ln t)^{a/(2+2a)}     & d=2\\
t^{1/(2+a)}                                  & d=1
\end{cases}
\end{equation}
Finally plugging \eqref{c_steady_non} into \eqref{estimate} and using \eqref{front_non} we obtain 
\begin{equation}
\label{total_non}
N(t)\sim 
\begin{cases}
t                               & d>d^c\\
t(\ln t)^{-1-a}            & d=d^c\\
t^{(d-2)/(2+2a)}        & 2<d<d^c\\
(\ln t)^2                    & d=2\\
\ln t                          & d=1
\end{cases}
\end{equation}
In particular, in the most relevant case of $d=3$ the total number of particles grows according to
\begin{equation}
N(t)\sim t^{1/(2+2a)} 
\end{equation}
Note also that for $d=1,2$ the growth of the total number of particles is very slow, namely logarithmic; interestingly, the asymptotic growth laws for the total number of particles are independent on $a$ in one and two dimensions. 

\subsection{Examples}

The simplest example of the concentration-dependent diffusion is $D=c$. This is relevant e.g. for the heat transfer in the so-called Maxwell gas \cite{M67,K70,TM80} since the coefficient of the thermal conductivity (and generally transport coefficients in the Maxwell gas) is proportional to temperature. In this case \eqref{front_non} shows that in the physically interesting dimensions $d=1,2,3$ the front propagates as 
\begin{equation}
\label{front_Maxwell}
\mathcal{R}(t)\sim 
\begin{cases}
t^{1/4}                        & d=3\\
t^{1/4} (\ln t)^{1/4}     & d=2\\
t^{1/3}                       & d=1
\end{cases}
\end{equation}

Another interesting example corresponds to $D=\sqrt{c}$. (For the hard-sphere gas, the transport coefficients are proportional to the square root of the temperature.) In this case the front propagates as 
\begin{equation}
\label{front_hard}
\mathcal{R}(t)\sim 
\begin{cases}
t^{1/3}                        & d=3\\
t^{1/3} (\ln t)^{1/6}     & d=2\\
t^{2/5}                       & d=1
\end{cases}
\end{equation}

\subsection{Exact Solution of the Mean-Field Equations in One Dimension}

Reaction-diffusion equations \eqref{RD_simple}, \eqref{RD_source}, and \eqref{non_source} for the single-species annihilation are mean-field in nature. These equations presumably provide asymptotically exact descriptions in three dimensions. In two dimensions, the mean-field treatment of the annihilation process is slightly wrong (the discrepancy is logarithmic),  and in one dimension the mean-field approach is flawed. Even when reaction terms are chosen to assure the correct qualitative behaviors, there is no closed equation for the density when the spatial dimension is below critical. In one dimension an exact description is feasible only in the case of a constant diffusion coefficient \cite{CRL89,K93}. It seems impossible to generalize these results to the case of the concentration-dependent diffusion. 

Nevertheless, it is methodologically interesting to solve the corrected mean-field equation, namely the one with the reaction term taken from Eq.~\eqref{RD_non} at $d=1$, in the one-dimensional setting. Thus we want to solve an equation
\begin{equation}
\label{non_source_1d}
\frac{\partial c}{\partial t} = \frac{\partial }{\partial x}\left(c^a\,\frac{\partial c}{\partial x}\right) - c^{3+a} + J\,\delta(x)\,\theta(t)
\end{equation}
This is a rather complicated initial-value problem. We limit ourselves to the steady-state solution. The governing ordinary differential equation 
\begin{equation}
\label{steady_1d}
\frac{d}{dx}\left(c^a\,\frac{d c}{d x}\right) - c^{3+a} + J\,\delta(x) = 0
\end{equation}
is solvable. Equation $(c^a c')' = c^{3+a}$, where the prime denotes the derivative with respect to $x$, admits an exact solution $c=A/x$ with $A = \sqrt{d^c}=\sqrt{4+2a}$. The translational invariance of equation $(c^a c')' = c^{3+a}$ implies that its general solution is given by $c=A/(x+x_0)$, where $x_0$ is an arbitrary constant. Returning to \eqref{steady_1d} and invoking the $x\leftrightarrow -x$ symmetry we see that the solution must be
\begin{equation}
\label{steady_1d_sol}
c = \frac{A}{|x|+x_0}\,, \qquad x_0 = \left(\frac{2A^{1+a}}{J}\right)^{1/(2+a)}
\end{equation}
The constant $x_0$ in \eqref{steady_1d_sol} was fixed by integrating \eqref{steady_1d} over a small region around the origin which gives
\begin{equation*}
c^a\,\frac{d c}{d x}\Big|_{x=-0}^{x=+0} = -  J 
\end{equation*}
and allows to express $x_0$ via the strength of the source. 

Solutions similar to \eqref{steady_1d_sol} have been obtained in \cite{book,CRL89}. A general case with the reaction term $c^n$ has been recently investigated in Ref.~\cite{BLL11} where it was additionally demonstrated that some of these solutions fit the experimental data \cite{HYBL} for morphogen gradient formation. (Morphogen gradients play a crucial role in developmental biology \cite{Crick,GB01}, e.g., they appear to be precursors to cell differentiation \cite{ERSB,Katja}.)

\section{Dissolution Process}
\label{DP}

Here we consider another reaction-diffusion process, a dissolution process. This process involves three distinct species. One species is composed of diffusing particles, while the particles constituting two other species are immobile. Diffusing atoms (species $A$) are injected into a small localized region of a $d-$dimensional lattice. The entire lattice is initially occupied by immobile atoms (species $B$), one $B$ atom per lattice site. Whenever an $A$ atom hops to a lattice site occupied by a $B$ atom, two atoms react to form an inert stable molecule (species $B^*$). This dissolution process is described by the reaction scheme 
\begin{equation}
\label{reaction_DP}
A(\text{diffusing})+B(\text{substrate})\rightarrow B^*(\text{stable})
\end{equation}
This model mimics a number of important industrial chemical processes including the dissolution of solids \cite{D87}, electropolishing \cite{EP}, corrosion and etching \cite{Meakin,Lar,KM91}, and erosion \cite{SBG}. Here we consider the dissolution process with a concentration-dependent diffusion coefficient. 

\begin{figure}
\centerline{
\includegraphics[width=10cm]{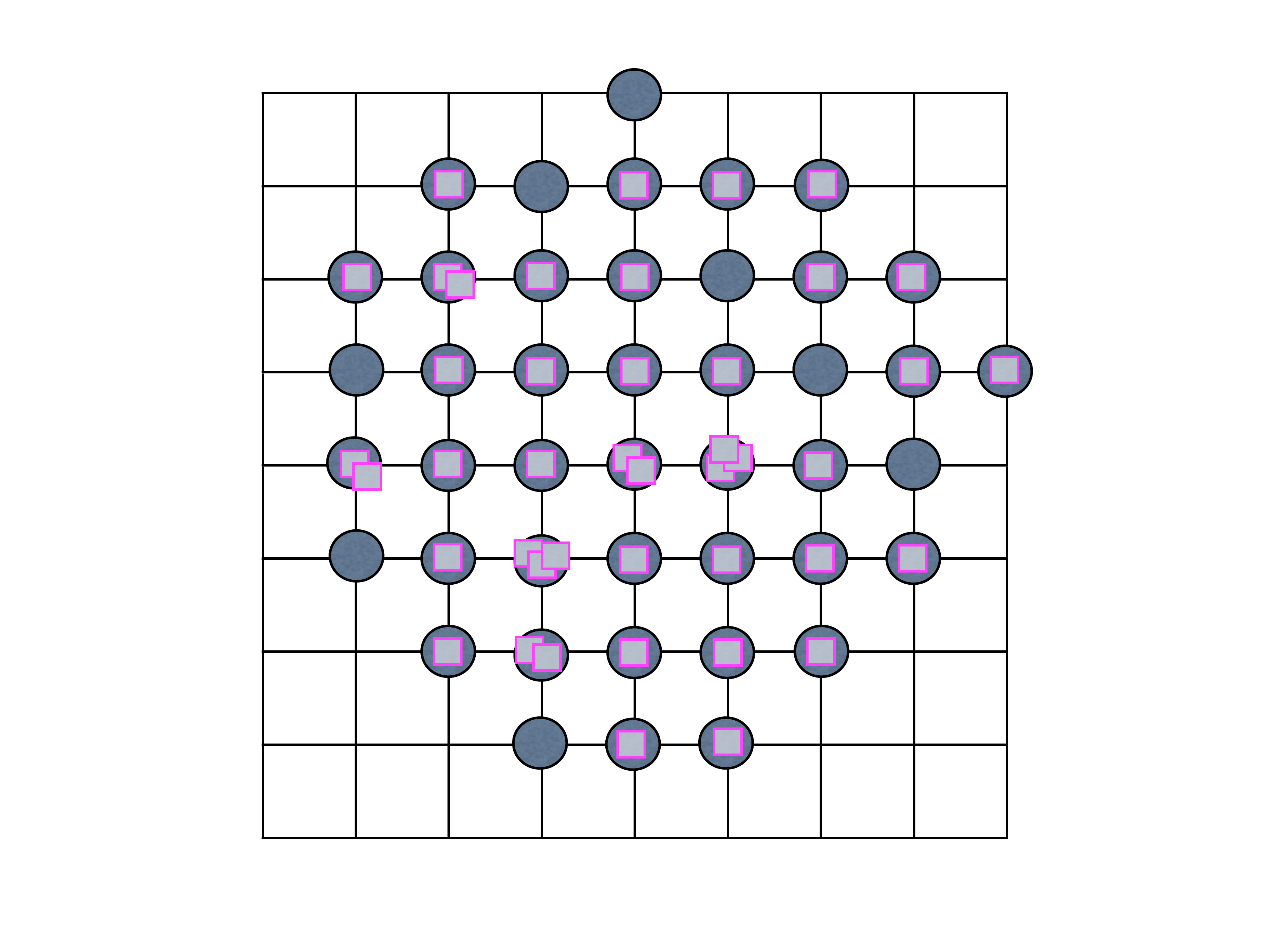}}
\caption{A growing droplet of $B^*$ molecules depicted as filled circles. The droplet is still rather small and hence the deviations from the round shape are visible. Each site with a $B^*$ molecule can contain $A$ atoms (depicted as squares) diffusing on the lattice and reacting with $B$ atoms (not displayed) which occupy sites outside the droplet.}
\label{DP_fig}
\end{figure}

The reaction \eqref{reaction_DP} proceeds at a certain finite rate. In many applications this rate greatly exceeds the hopping rate. Thus we shall assume that the reaction proceeds instantaneously, so each site contains either a $B$ atom or a $B^*$ molecule. As the process develops, the system can be separated into two parts: The droplet that contains no $B$ atoms (every lattice site inside the droplet is occupied by a $B^*$ molecule and can contain an arbitrary number of $A$ atoms) and the rest of the system that contains only $B$ atoms. 

The droplet is a growing random set (Fig.~\ref{DP_fig}). In the large time limit, the droplet becomes relatively closer to the ball. This is not obvious, yet it has been proven in the case of constant diffusivity \cite{BGL,Stefan}. (Even the magnitude of the fluctuations in this situation has been recently established, see \cite{IDLA_1,IDLA_2} and references therein.) In the following we assume that the same qualitative behavior continues to hold in the concentration-dependent case. The droplet is therefore asymptotically a ball of radius $R(t)$ which is determined in the process of solving the problem. We ignore the fluctuations, so the results should be applicable for a sufficiently large time. 

The concentration $c({\bf r},t)$ of $A$ atoms (that is, the average number of $A$ atoms per lattice site) satisfies a non-linear diffusion equation 
\begin{equation}
\label{RD:c2d}
\frac{\partial c}{\partial t} = D\left(\frac{\partial^2}{\partial r^2} + 
\frac{d-1}{r}\,\frac{\partial}{\partial r}\right)\frac{c^{1+a}}{1+a} + J\delta({\bf r})\,\theta(t)
\end{equation}
inside the droplet $0\leq r\leq R(t)$. The diffusion equation \eqref{RD:c2d} should be supplemented by the adsorbing condition
\begin{equation}
\label{BC:c}
c(r=R(t), t>0)=0
\end{equation}
and the Stefan boundary condition
\begin{equation}
\label{Stefan}
\frac{dR}{dt} = - Dc^a\,\frac{\partial c}{\partial r}\Big|_{r=R} 
\end{equation}
The boundary moves and its position has to be determined in the process of solution. Therefore mathematically we arrive at the Stefan problem \cite{CJ59,C87}. We analyze \eqref{RD:c2d}--\eqref{Stefan} using the same approach \cite{Lar,K} as in the case of the concentration-independent diffusion coefficient. 

The original process occurs on the lattice and therefore we set the lattice spacing to unity; this implies that the spatial coordinates ${\bf r}$, the droplet radius $R(t)$, and the concentration $c({\bf r},t)$ are all dimensionless quantities, while the hopping rate $D$ and the strength of the source have the dimension of inverse time: $[D]=[J]=1/(\text{time})$. The ratio $J/D$, the dimensionless flux, plays an important role in the problem. 

We limit ourselves with the two-dimensional case which is most important in applications \cite{D87,EP,SBG}. Seeking a solution in the scaling form
\begin{equation}
\label{scaling}
c(r,t) = c(\xi), \quad \xi=\frac{r}{R}
\end{equation}
we reduce the diffusion equation \eqref{RD:c2d} to
\begin{equation}
\label{RD:c2d-scal}
c'' + \frac{1}{\xi}\,c' + \frac{a}{c}\,(c')^2= - \frac{R\dot R}{D}\,\frac{\xi c'}{c^a}
\end{equation}
where $(\cdot)'\equiv d(\cdot)/d\xi$ and $\dot R=dR/dt$. Equation \eqref{RD:c2d-scal}  is consistent if $R\dot R/D$, which is in principle a function of time, is actually a constant. Hence we write
\begin{equation}
\label{radius:2d}
R^2 = 2\beta Dt,
\end{equation}
and recast \eqref{RD:c2d-scal} into 
\begin{equation}
\label{c2d-scal}
c'' + \xi^{-1} c' + \frac{a}{c}\,(c')^2 + \frac{\beta \xi c'}{c^a} = 0
\end{equation}
which must be solved subject to 
\begin{equation}
\label{BC}
c=0, \quad c^ac' +\beta=0 \qquad \text{when}\quad \xi=1
\end{equation}
Equation \eqref{BC} implies that $c\simeq [\beta(1+a)(1-\xi)]^{1/(1+a)}$ in the $\xi\to 1$ limit.
The conservation of the total number of $A$ atoms gives
\begin{equation}
\label{conserv_law:2d}
Jt = \int_0^R c(r,t)\,2\pi r\,dr + \pi R^2
\end{equation}
Using Eqs.~\eqref{scaling} and \eqref{radius:2d}, we re-write \eqref{conserv_law:2d} in the form
\begin{equation}
\label{flux_beta}
\frac{J}{4\pi D \beta} = \int_0^1 d\xi\,\xi c(\xi) + \frac{1}{2}
\end{equation}
which implicitly determines $\beta$ in terms of the dimensionless flux $J/D$. The right-hand side of \eqref{flux_beta} depends, of course, on $\beta$ since the concentration $c(\xi)$ is determined by solving Eqs.~\eqref{c2d-scal}--\eqref{BC} which contain $\beta$.

\section{Summary}
\label{Sum}

In annihilation processes the concentration decays to zero, and even in annihilation processes driven by a localized source the concentration asymptotically vanishes when the distance from the source increases to infinity. Hence it is vitally important to know the behavior of $D(c)$ in the $c\to 0$ limit. If $D(0)=0$, the behaviors qualitatively differ from the classical case of constant diffusivity. 

We studied single-species and two-species diffusion-controlled annihilation processes in the situation when the diffusion coefficient vanishes algebraically, $D\sim c^a$, in the $c\to 0$ limit. In the homogeneous case, the critical dimensions are $d_c=2$ (for the single-species annihilation) and $d_c=4$ (for the two-species annihilation), and in the physically interesting dimensions the long-time asymptotics are, respectively, given by \eqref{DLA_non} and \eqref{DLA_2}. For the two-species annihilation we also investigated the width of the reaction zones, both in the homogeneous setting (the width of the zone between adjacent domains) and the inhomogeneous setting (when initially the reactants occupy complimentary half spaces); the chief results are given by Eqs.~\eqref{lengths_1d}--\eqref{lengths_2d} and \eqref{width}. We then studied the single-species annihilation process in the situation when the spatially localized source drives the system toward the non-equilibrium steady state. We showed that two critical dimensions demarcate different behaviors: $d_c=2$ which coincides with the critical dimension of the homogeneous process, and $d^c=4+2a$ which depends on the exponent $a$. Our main results are given by \eqref{c_steady_non}--\eqref{total_non}. 

We also investigated the dissolution process involving diffusing atoms injected into a localized region, an immobile species initially fully occupying the lattice, and another immobile species formed as the reaction product between them, Eq.~\eqref{reaction_DP}. An asymptotically spherical growing droplet that contains no original immobile atoms is formed as the process develops. Mathematically, one needs to solve the Stefan problem as the radius of the droplet is determined in the process of solution. In the most important in applications two-dimensional setting, the solution is self-similar and this allowed us to reduce the problem to an ordinary differential equation. This equation is analytically soluble only in the case of concentration-independent diffusivity when the governing equation is linear \cite{Lar}. In this classical case, the probability that a diffusing atom has not become a part of a molecule has been recently determined \cite{K}. It appears possible to generalize the analysis of Ref.~\cite{K} and probe the first passage characteristics in the case of concentration-dependent diffusivity. 

Finally we mention recent studies \cite{HK,PLH} of fluctuations in diffusion processes with concentration-dependent diffusivity. An interesting challenge is to probe the role of fluctuations for annihilation processes with concentration-dependent diffusivity.


\begin{thebibliography}{99}
\bibitem{diff_caveat}
      The diffusive growth of the depletion zone [Eq.~\eqref{cF:scaling}] is the consequence of the
      non-vanishing initial condition \eqref{excess:IC}; only for the spread into a vacuum does the size of the
      perturbed region exhibit a sub-diffusive growth, $\ell\sim t^{1/(2+a)}$ in one
      dimension, dependent on the mobility exponent $a$.
\end{thebibliography}
\end{document}